\documentclass[12pt]{emulateapj}

\def\etl{et al.}
\def\nii{[\ion{N}{2}]}
\def\oii{[\ion{O}{2}]}
\def\oiii{[\ion{O}{3}]}

\def\oiilam{[\ion{O}{2}]~$\lambda3727$}
\def\oiiilam{[\ion{O}{3}]~$\lambda5007$}
\def\oiidoublet{[\ion{O}{2}]~$\lambda\lambda3726,3729$}
\def\oiim{[O II]}
\def\ha{H$\alpha$}
\def\hb{H$\beta$}
\def\ergs{\mathrm{erg~s^{-1}}}

\newcommand{\loii}{\ensuremath{L_{\mathrm{\oiim}}}}
\newcommand{\loiistar}{\ensuremath{L_{\mathrm{\oiim}}^{*}}}

\usepackage{color}
\usepackage{ulem}
\usepackage{lscape}

\begin{document}
\shorttitle{\oii~LF at $z\sim1$}
\shortauthors{Zhu \etl}
\title {
The \oiilam~Luminosity Function at $z\sim1$}

\author{
 Guangtun Zhu\altaffilmark{1},
 John Moustakas\altaffilmark{1}, and
 Michael R. Blanton\altaffilmark{1}
} 
\altaffiltext{1}{
Center for Cosmology and Particle Physics, Department of Physics, New York University,
4 Washington Place, New York, NY 10003, gz323@nyu.edu, john.moustakas@nyu.edu, michael.blanton@nyu.edu}
%\received{---------------}
%\accepted{---------------}

\begin{abstract}
  We measure the evolution of the \oiilam{} luminosity function at
  $0.75<z<1.45$ using high-resolution spectroscopy of $\sim14,000$
  galaxies observed by the DEEP2 galaxy redshift survey.  We find that
  brighter than $L_{\mathrm{\oiim}}=10^{42}~\ergs$ the luminosity
  function is well-represented by a power law $dN/dL \propto
  L^{\alpha}$ with slope $\alpha \sim -3$.  The number density of
  \oii-emitting galaxies above this luminosity declines by a factor of
  $\gtrsim2.5$ between $z \sim 1.35$ and $z \sim 0.84$.  In the limit
  of no number-density evolution, the characteristic \oii{}
  luminosity, \loiistar, defined as the luminosity where the
  space density equals $10^{-3.5}~\mathrm{dex^{-1}~Mpc^{-3}} $,
  declines by a factor of $\sim1.8$ over the same redshift interval.
  Assuming that \loii{} is proportional to the star-formation
    rate (SFR), and negligible change in the typical dust attenuation
    in galaxies at fixed \oii{} luminosity, the measured decline in
    \loiistar{} implies a $\sim25\%$ per Gyr decrease in the amount of
    star formation in galaxies during this epoch.
  Adopting a faint-end power-law slope of $-1.3\pm0.2$, we
    derive the comoving SFR density in four redshift bins centered
    around $z\sim1$ by integrating the observed \oii~luminosity
    function using a local, empirical calibration between \loii{} and
    SFR, which statistically accounts for variations in dust
    attenuation and metallicity among galaxies. We find that
  our estimate of the SFR density at $z\sim1$ is consistent
  with previous measurements based on a variety of independent
  SFR indicators.
\end{abstract}

\keywords{galaxies: evolution  --- galaxies: luminosity function ---
  stars: formation} 

\section {Introduction} 
Measuring the comoving space density of the star formation rate (SFR) as a function of cosmic
epoch is one of the key issues concerning the study of galaxy formation and evolution.
The current picture is that the star formation peaked at $z\sim1-3$, and then declined 
by roughly an order-of-magnitude to the present day
(see, e.g., \citealt{madau96, lilly96, hopkins04, hopkins06}).

Among the most direct indicators of the instantaneous SFR in galaxies
is the \ha~$\lambda6563$ recombination line (\citealt{kennicutt83,
  kennicutt92}).  \ha~can be observed in the optical in the local
Universe (e.g., \citealt{gallego95, tresse98, kennicutt08}).  However,
at $z \gtrsim 0.4$, it must be observed in the near infrared (e.g.,
\citealt{yan99, glazebrook99, hopkins00, tresse02}), or other less
direct SFR indicators such as the \oiidoublet{} doublet must be used.

Compared to \ha, \oii{} is only indirectly coupled to the ionizing
continuum from massive stars, and is more sensitive to variations in
metal abundance, excitation, and dust attenuation.  Nevertheless,
because of its intrinsic strength and blue rest-frame wavelength,
\oii~remains a good alternative SFR indicator for high-redshift
galaxies (\citealt{kennicutt98, jansen01, kewley04, mouhcine05,
  moustakas06a}).  In the past decade, a number of investigators have
measured the star formation rate density, $\rho_{\rm SFR}$, at high
redshift by studying the \oii~luminosity function (LF), either using
spectroscopy (e.g.,
\citealt{hammer97,hogg98,gallego02,teplitz03,rigopoulou05}), or
narrow-band imaging (e.g., \citealt{hippelein03};
\citealt{ly07}, hereafter Ly07; \citealt{takahashi07},
  hereafter Takahashi07).
Unfortunately, these studies have been hampered by small sample size,
small volume probed, and an inconsistent treatment of dust
obscuration. Methods like narrow-band imaging also suffer from
difficulties in continuum subtraction and contamination from other
emission lines.

To circumvent these issues, we measure the \oii{} luminosity
  function at $z\sim1$ using data from the Deep Extragalactic
  Evolutionary Probe 2 survey (DEEP2\footnote{\tt
    http://deep.berkeley.edu}; \citealt{davis03}).  DEEP2 has obtained
  high-resolution spectra for $\sim 50,000$ objects over four separate
  fields, making it the largest existing spectroscopic redshift survey
  of galaxies at these redshifts.  We use these data to measure the
  \oii{} luminosity function in four redshift bins at $0.75<z<1.45$.

In $\S$2, we briefly describe our sample and the method used to
calculate the LF. In \S3 and \S4 we present the observed \oii{} LF and
its evolution with redshift, respectively.  Finally, in $\S$5, we
compute $\rho_{\rm SFR}$ in several redshift bins centered around 
$z\sim1$, and we summarize our principal conclusions in $\S$6. 

Throughout this work, we adopt a $\Lambda \mathrm{CDM}$ cosmology with
$\Omega_{\mathrm{m}}=0.3$, $\Omega_\Lambda=0.7$ and
$H_0=70~\mathrm{km~s^{-1}~Mpc^{-1}}$.  All magnitudes are on the AB
system.

\section{Data and Method}
\subsection{Data: DEEP2 DR3}

We select our sample from the DEEP2 third public data release (DR3),
which includes $BRI$ photometry and spectra for $\sim50,000$ galaxies
in four widely separated fields.  The high-resolution ($R\equiv
\lambda/\delta\lambda \sim5000$) spectra, which were acquired using
the Keck-II/DEIMOS spectrograph \citep{faber03}, span $6525 - 9120$
\AA.  We thus restrict our analysis to the redshift range $0.75 < z <
1.45$, where \oii~is measurable. To study the evolution of the \oii~
LF, we further split each bin into four additional redshift bins:
$0.75 < z < 0.93$; $0.93 < z < 1.10$; $1.10 < z < 1.28$; and $1.28 < z <
1.45$.
We refer the reader to \citet{davis03},
\citet{coil04}, and \citet{davis07} for additional details regarding
the DEEP2 survey.

In addition to the flux cut, $R<24.1$~mag, the DEEP2 team applied the
following color cuts to preselect galaxies at $z > 0.7$ in Fields 2,
3, and 4:
\begin{eqnarray}
B-R &<& 2.35 (R-I) - 0.25\ {\rm or}; \nonumber \\
R-I &>& 1.15\ {\rm or}; \nonumber \\ 
B-R &<& 0.50.
\label{eq1}
\end{eqnarray}

\noindent In the Extended Groth Strip (EGS, Field 1), to test their
sample preselection method, the DEEP2 team did not apply the $BRI$
color cuts. However, in order to make all four fields consistent, here
we apply the color cuts given by equation~(\ref{eq1}) to the redshift
and photometric catalogs in the EGS field.  We apply additional
angular cuts to avoid survey edges and to exclude gaps in the
spectroscopy.  Our final catalog contains $36,118$ objects, of which
$24,729$ have accurate redshifts (quality $Q = 3$ or $4$ as defined by
\citealt{davis07}).  The areas of the four fields are $0.38$, $0.56$,
$0.88$ and $0.63~\mathrm{deg}^2$, respectively, and the total area is
$2.45~\mathrm{deg}^2$.

Next we describe how we derive the integrated \oii{} luminosity for
each galaxy.  Unfortunately, the DEEP2 spectra are not
flux-calibrated; therefore we infer the \oii{} luminosity by
multiplying the rest-frame emission-line equivalent width (EW)
measured in the optical spectra, by the continuum luminosity around
$3727$~\AA{} inferred from fitting the $BRI$ photometry.  This method
has the advantage that it is insensitive to the absolute calibration
of the optical spectra.  However, the $1\arcsec$ wide slit in the
DEEP2 survey may not enclose all the line-emitting regions in the
galaxy, thus this method does assume that the relative intensity of
star formation and stellar light inside and outside the slit does not
vary significantly.  Without spatially resolved spectroscopy, however,
this assumption is difficult to test directly.

To measure the emission-line EW, we model each component of the
\oiidoublet~doublet simultaneously using two Gaussian profiles,
constrained to have the same intrinsic velocity width and a fixed
wavelength separation, and use a smooth $B$-spline to estimate the
continuum level around \oii{}.  Dividing the total \oii{} flux by the
detected continuum yields the rest-frame EW(\oii{}) in \AA.  We then
fit the broad-band $BRI$ photometry at the known redshift to obtain the
best-fitting spectral energy distribution (SED) using the {\tt
  deep\_kcorrect} routine in {\tt kcorrect}\footnote{\tt
  http://cosmo.nyu.edu/blanton/kcorrect} (v4.1.4;
\citealt{blanton07}).  Finally, we multiply the EW by the flux-density
of the best-fitting SED at $3727$~\AA{} to obtain the integrated
\oii{} luminosity.  Because the effective wavelengths of all the $BRI$
filter bandpasses lie blueward of $3700$~\AA{} above $z\sim1.2$, this
technique does require extrapolating the best-fitting SED beyond the
effective $I$-band wavelength to estimate the $3727$~\AA{} continuum
luminosity for the galaxies in our highest redshift bin; however, the
uncertainty introduced by this extrapolation is negligible.

Our \oii{} measurement technique is similar to that used by the DEEP2
team (\citealt{weiner07, cooper08}), although the procedures used to
compute $K$-corrections are totally independent.  A comparison of our
measurements shows no systematic differences and a $\sim25\%$ scatter,
which is comparable to the typical measurement error.  We also checked
our work by replacing our measurements with theirs, and obtained
consistent results.  In the following analysis we use our
measurements. 

We consider an \oii{} measurement with signal to noise ratio ($S/N$)
larger than $5$ as reliable.  In detail, our results are not sensitive
to the specific $S/N$ cut used since it only affects weak
\oii~detections, for which we are incomplete anyway.  For example,
using $S/N>2$ has no significant effect on our conclusions.  Our final
sample of \oii-emitting galaxies contains $13,944$ objects.

\subsection{Method: $1/V_{\mathrm{max}}$ Method}

To calculate the luminosity function, we use the non-parametric
$1/V_{\mathrm{max}}$ method \citep{felten76}.  For a given galaxy, we
calculate:
\begin{equation}
V_{\mathrm{max}} = \frac{1}{3} \int d\Omega \int_{z_{\mathrm{min}}}^{z_{\mathrm{max}}} dz
\frac{d[D_c(z)^3]}{dz} f(z) \mathrm{,}
\end{equation}
\noindent which is suitable for a spatially flat universe.  The
angular integral is limited to the DEEP2 area, $D_c(z)$ is the
comoving distance as defined by \citet{hogg99}, and $f(z)$ expresses
the probability of selecting each galaxy in our sample.  We assume
that $f(z)$ is given by the product of four quantities:
$f=f_{\mathrm{target}}\times f_{\mathrm{success}}\times
f_{\mathrm{cover}}\times f_{\mathrm{cut}}$, where
$f_{\mathrm{target}}$ is the rate at which a source of a given
$R$-band magnitude and $B-R$ and $R-I$ color was targeted [see
  eq.~(\ref{eq1})], $f_{\mathrm{success}}$ is the rate at which a
redshift was successfully measured, $f_{\mathrm{cover}}$ is the
fraction of DEEP2 spectra where the wavelength coverage includes the
redshifted \oii{} line,
and $f_{\mathrm{cut}}$ is a step function, which we determined using
the $BRI$ magnitudes and the \oii{} $S/N$, as described below.  We
experimented with including the galaxy surface brightness as a fifth
variable in the completeness function (see, e.g., \citealt{lin08}),
but found that it did not significantly affect our measured LFs.

For a given galaxy, we calculate $V_{\mathrm{max}}$ using a Monte
Carlo method (see \citealt{blanton06}).  We randomly choose 1200
values of redshift $z$ between $z_{\mathrm{min}}$ and
$z_{\mathrm{max}}$, uniformly distributed in volume.  For each
redshift, we calculate what the magnitudes of the object would be
using {\tt deep\_kcorrect}. To estimate the \oii{} $S/N$ at each mock
redshift, we first determine the mean noise spectrum for every DEEP2
mask.  Then, for a given galaxy, we randomly choose a mask and the
corresponding noise spectrum and compare the noise at the actual
observed wavelength of \oii{} to that at the simulated wavelength as
if the galaxy were at the simulated redshift, and obtain the new
$S/N$.  We then apply the appropriate flux cut, color cuts, \oii{}
$S/N$ cut and the completeness function to determine the fraction of
mock sources that would have passed our selection criteria.  Finally,
$V_{\mathrm{max}}$ is given by the comoving volume multiplied by this
fraction.

When determining the completeness function, we assume a $100\%$
redshift success rate for blue galaxies; that is, we assume that the
only targeted blue galaxies without successfully measured redshifts
are those for which \oii~ falls outside the wavelength range of the
spectra, and thus outside the redshift range studied here.  Here, we
define blue galaxies using the same definition as \citet{willmer06},
who also made the same assumption regarding blue galaxies without
well-measured redshifts.  This assumption is especially reasonable in
our analysis because a well-detected \oii{} doublet will always result
in a well-measured redshift.  Finally, we tested our completeness
function by calculating the $B$-band LF for all galaxies and comparing
with \citet{willmer06}, and found excellent statistical agreement, and
no systematic differences.

\section{The Observed \oii~Luminosity Function}

%=======================
% Figure 1:
%=======================
\begin{figure}
\epsscale{1.25}
\plotone{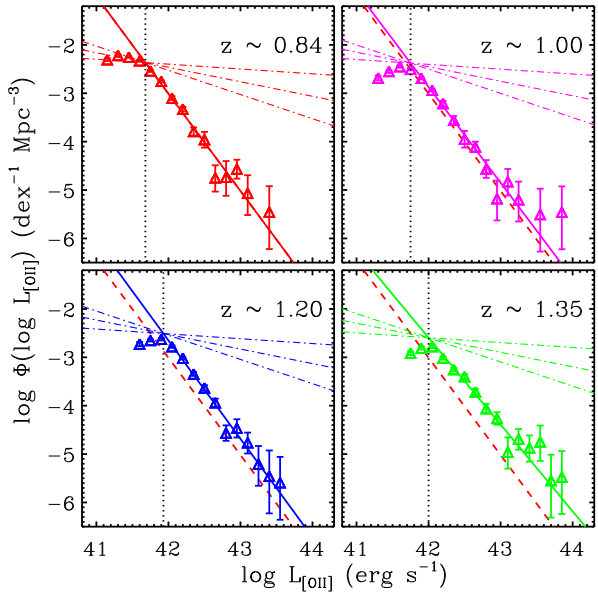}
\caption{Observed \oii~luminosity function in four redshift bins. The
  solid lines are the power-law fits to the bright part of the
  luminosity function. The dashed lines in the three highest redshift
  bins correspond to the solid line in the top-left panel.  The three
  dashed-dotted lines correspond to three different faint-end slopes,
  $-1.1$, $-1.3$ and $-1.5$, and the dotted lines in each panel
  indicate the {\it turnover} in the luminosity function.}
\label{fig1}
\end{figure}

We measure the \oii~luminosity function in four redshift bins: $0.75 <
z < 0.93$; $0.93 < z < 1.10$; $1.10 < z < 1.28$; and $1.28 < z <
1.45$, according to where the \oii{} doublet could be measured
reliably in the DEEP2 spectra.  The median redshifts of the
\oii-emitting galaxies within these four redshift bins are $0.84$,
$1.01$, $1.19$, and $1.35$, containing $5471$, $3771$, $3018$ and
$1684$ galaxies in each bin, respectively.

The resulting \oii~LFs are tabulated in Table~\ref{tbl1} and
illustrated in Figure~\ref{fig1}.  The error bars shown in the figure
are $84.13\%$ confidence Poisson upper limits and lower limits
estimated using the approximate formulas given by \citet{gehrels86}
[see eqs. (10) and (14) in that paper]. The distribution of the LF
must follow a scaled Poisson distribution.  To determine the scaling
factor, within each luminosity bin, we define the effective weight
($W_{\mathrm{eff}}$) by:

\begin{equation}
W_{\mathrm{eff}} =
\left[\sum_{i}{\frac{1}{(V_\mathrm{max})^2_i}}\right]\left/\left[{\sum_{i}{\frac{1}{(V_\mathrm{max})_i}}}\right]\right.~\mathrm{,}
\end{equation}
\noindent and the effective number ($N_{\mathrm{eff}}$) of objects by:
\begin{equation}
N_{\mathrm{eff}} = \left[\sum_{i}{\frac{1}{(V_\mathrm{max})_i}}\right]\left/{W_{\mathrm{eff}}}\right.~\mathrm{.}
\end{equation}

\noindent We then calculate the upper limit and lower limit for the
effective number, and multiply them by the effective weight to obtain
the upper limit and lower limit for the luminosity density in each
bin.  For comparison, the square root Poisson error gives:

\begin{equation}
\sigma_{\Phi} = \sqrt{N_{\mathrm{eff}}}W_{\mathrm{eff}} =
\sqrt{\sum_{i} \frac{1}{(V_{\mathrm{max}})^2_i}}~\mathrm{,}
\end{equation}

\noindent which is commonly used in the literature.  The Poisson
errors, however, do not include the effects of cosmic variance.
Because we have four widely separated fields, we determine the error
due to cosmic variance, $\sigma_{\mathrm{cv}}$, by calculating the
variance among the four independent fields, and list the results in
Table~\ref{tbl1}.

Examining Figure~\ref{fig1}, the commonly used \citet{schechter76}
function is clearly a poor representation of the data.  Instead, we
model the observed \oii~LF in each redshift bin as a power law:

\begin{equation}
\Phi(\log L) d(\log L) = 10^{(\alpha+1)(\log L - 42.5) + \beta}d(\log
L) \mathrm{,}
\end{equation}

\noindent where $L$ is $L_{\mathrm{\oiim}}$ in $\ergs$, and $\alpha$
and $\beta$ are dimensionless parameters.  We find the best fitting
parameters ($\alpha, \beta$) using a non-linear least square fit to
the \oii~LFs weighted by the average Poisson errors. These parameters
are presented in Table~\ref{tbl2}.  Our results show that the bright
part of each LF can be represented by a power law $dN/dL \propto
L^{\alpha}$ with slope $\alpha \sim -3$. The slope for the highest
redshift bin is the flattest: $-2.79\pm0.12$. 
However, it is possible that the slope in this bin 
may be underestimated due to incompleteness (see below).

%========================
% Figure 2:
%========================
\begin{figure}
\epsscale{1.2}
\plotone{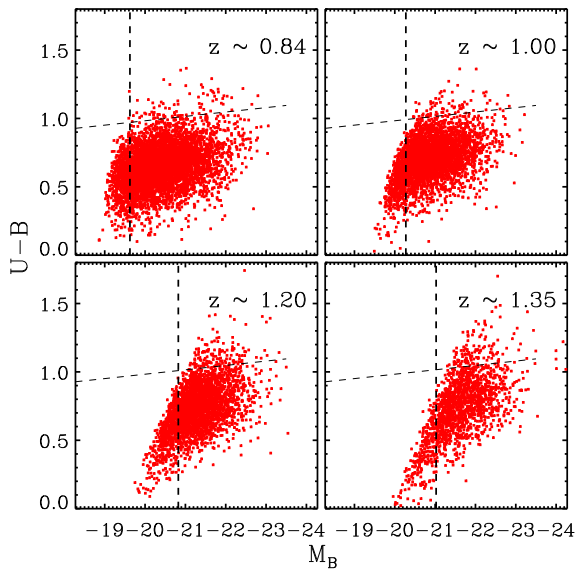}
\caption{Rest-frame color-magnitude diagrams.  The thin dashed lines
  show the division between blue galaxies and red galaxies (see
  \citealt{willmer06}), and the thick dashed lines show the
  approximate $M_B$ completeness limit for our sample.}
\label{fig2}
\end{figure}

\begin{figure}
\epsscale{1.2}
\plotone{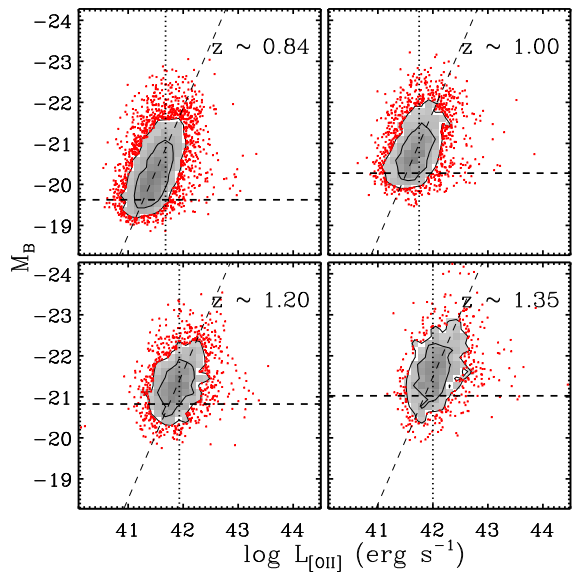}
\caption{$M_B-\log(L_\mathrm{\oiim})$ distribution.  The dotted lines
  are the same as in Figure~\ref{fig1}. The thick dashed lines show
  the $M_B$ approximate completeness limit for blue galaxies (see
  Figure~\ref{fig2}). The thin dashed lines have a slope of
  $-3.1~\mathrm{mag~dex^{-1}}$, and show the approximate median
  relation between $M_B$ and $\log(L_\mathrm{\oiim})$.}
\label{fig3}
\end{figure}

Unfortunately, we are unable to constrain the faint end of the
\oii~LF, especially in the highest redshift bin.  Because we do not
know the intrinsic number of \oii{} emitters within a certain \oii{}
luminosity bin, we can not calculate the completeness of \oii{}
luminosity directly, but can only infer that from the completeness of
broad-band properties.  To analyze the completeness, and in particular
to test the significance of the observed {\it turnover} (TO) in the LF
({\it vertical dotted lines}; Fig.~\ref{fig1}), we construct two
diagnostic diagrams.  In Figure~\ref{fig2} we show the $U-B$ versus
$M_{B}$ color-magnitude diagram for our sample.  We plot the
approximate $B$-band completeness limits for blue galaxies as dashed
lines, where the sloping color cut has been defined by
\citet{willmer06}, and the thick vertical dashed line roughly
corresponds to where the data and the color cut begin to deviate.  In
Figure~\ref{fig3} we plot the density distribution of points in the
$M_B - \log(L_{\mathrm{\oiim}})$ plane in grey scale; the two contours
enclose $50\%$ and $80\%$ of the points, respectively.  The thin
dashed lines in this figure all have a slope of
$-3.1~\mathrm{mag~dex^{-1}}$.  This slope is formally consistent with
performing a mean ordinary least square fit to the data
(\citealt{isobe90}).  For our purposes, we note that the line roughly
bisects the distribution of points in each panel, i.e., it
approximates the median relation between \loii{} and $M_{B}$.  The
thick horizontal dashed lines in each panel are equivalent to the
thick vertical dashed line plotted in the respective panels in
Figure~\ref{fig2}, and the vertical dotted lines give the position of
the turnover in the respective \oii~LF (Fig.~\ref{fig1}).  The
majority of the galaxies missing from our sample should be along and
to the left of the thin dashed line.

Figure~\ref{fig3} demonstrates that, brighter than the turnover, we
expect the sample to be complete because the majority of the
unobserved galaxies below the $M_B$ completeness limit are along and
to the left of the thin dashed line.  Fainter than the turnover, the
sample becomes increasingly incomplete.  In the two lowest redshift
bins, the turnovers appear to be significant, because fainter than the
turnover the difference between the measured LF and the extrapolation
of the power law fitted to the bright part of the LF is so significant
that it is unlikely to be due to incompleteness.  In the two highest
redshift bins, however, because the $M_B$ completeness limit is very
bright, it is possible that the turnovers are artificial, caused by
the incompleteness of the survey.  In the highest redshift bin, the
$M_B$ completeness limit is so bright that the slope of the LF may be
even steeper than we have derived.

%========================
% Figure 3:
%========================

To summarize, we are unable to constrain the faint end of the
\oii~luminosity function.  However, we emphasize that the evolutionary
analysis presented in $\S4$ is unaffected by our inability to
constrain the faint-end slope, because we restrict our analysis to the
bright part of the \oii{} LF where we are statistically complete.  In
$\S5$, when integrating the \oii{} LF to obtain an estimate of the SFR
density, we do make some simplified assumptions regarding the form of
the faint end of the LF.

\begin{figure}
\epsscale{1.2} \plotone{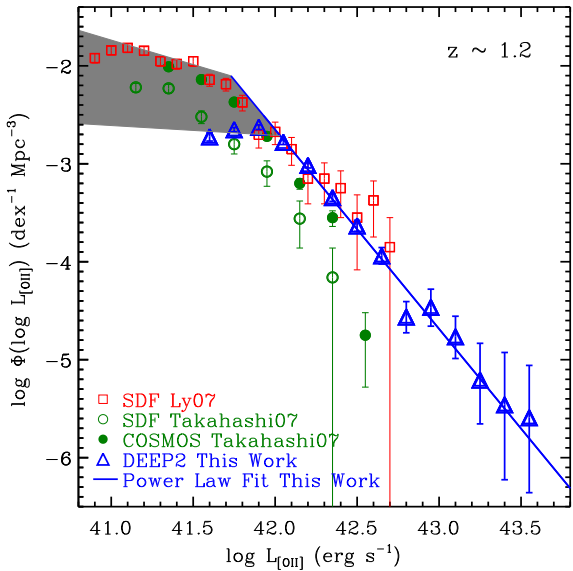}
\caption{Comparison of our \oii~luminosity function at $z\sim1.20$
  with other recent measurements from the literature based on
  narrow-band imaging.
The open triangles and the solid line
    represent our data and best-fitting power law (Fig.~\ref{fig1},
    {\it lower-left panel}). The open squares represent the LF in
  the Subaru Deep Field (SDF) calculated by \citet{ly07} at $z \sim 1.18$.
The open and filled circles correspond to observations of
  galaxies at $z \sim 1.19$ in the SDF and COSMOS fields, respectively
  \citep{takahashi07}.  Finally, the shaded region represents our
assumption for the faint-end slope when we convert the \oii{} LF to
SFR density in $\S5$.}
\label{fig4}
\end{figure}

In Figure~\ref{fig4}, we compare our results at $z\sim1.2$ with
three other recent coeval measurements of the \oii{} LF based
  on narrow-band observations (Ly07, Takahashi07).  Using Subaru/NB816
  narrow-band imaging of the $875~\mathrm{arcmin}^2$ Subaru Deep Field
  (SDF) by \citet{kashikawa04}, Ly07 and Takahashi07 identified $894$
  and $602$ \oii-emitting galaxies, respectively.  In the
  $2~\mathrm{deg}^2$ Cosmic Evolution Survey field (COSMOS;
  \citealt{scoville07}), Takahashi07 used the same narrow-band filter
  to identify $5824$ \oii-emitting galaxies at $z\sim1.2$
  \citep{taniguchi07}.  Over the range of \oii{} luminosity where all
  the surveys are complete, $10^{42.0}~\ergs \lesssim \loii \lesssim
  10^{42.7}~\ergs$, we find that our LF from DEEP2 agrees very well
  with the luminosity functions derived by Ly07 and Takahashi07 in the
  SDF and COSMOS fields, respectively.  It is not clear why the LF of
  the SDF derived by Takahashi07 is so discrepant with the other
  surveys, although cosmic variance may play a role.  Nevertheless,
  this comparison shows that: (1) our assumptions regarding the
  faint-end slope are reasonable; and (2) the bright end of the
  \oii~LF is clearly a power law, not a Schechter function.

%========================
% Figure 4:
%========================

\section{The Evolution of the \oii~Luminosity Function}

\begin{figure}
\epsscale{1.2}
\plotone{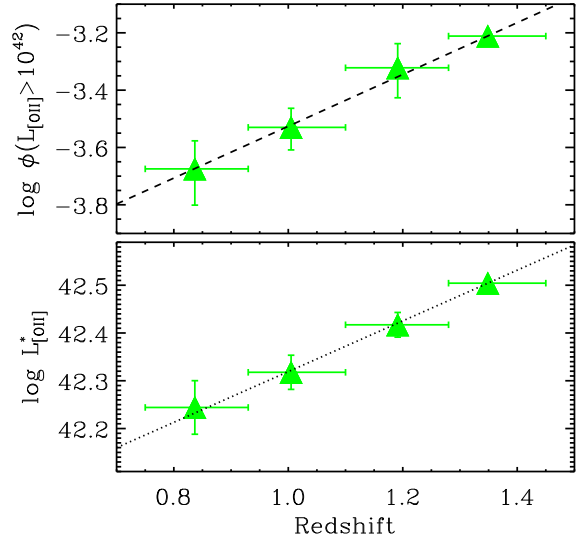}
\caption{Evolution of the \oii~luminosity function.  ({\it Top})
  Evolution of the total number density in $\mathrm{Mpc^{-3}}$ of the
  strongest \oii-emitting galaxies with
  $L_\mathrm{\oiim}>10^{42}~\ergs$.  The dashed line is the linear fit
  to the data given by eq.~(7).  ({\it Bottom}) Evolution of the
  characteristic luminosity, in $\ergs$, defined by where the space
  density of \oii-emitting galaxies equals
  $10^{-3.5}~\mathrm{dex^{-1}Mpc^{-3}}$. The dotted line is the linear
  fit to the data given by eq.~(8).}
\label{fig5}
\end{figure}

The uncertainty in the faint end of the \oii{} LF prevents us from
obtaining a reliable estimate of the total \oii~luminosity density.
Therefore, we focus instead on the integrated number density,
$\phi\equiv\int_{L}^{\infty}\Phi(L)\,dL$, of the strongest
\oii-emitting galaxies.  Assuming that the turnover luminosity in the
highest redshift bin is $\leq 10^{42}~\ergs$, we integrate our model
of the LF in each redshift bin over $L_{\mathrm{\oiim}} >
10^{42}~\ergs$, and present the results in the top panel of
Figure~\ref{fig5} and in Table~\ref{tbl2}.  The horizontal error bars
indicate the range of each redshift bin, and the vertical error bars
are given by the cosmic variance among the four fields, which dominate
the error budget.  We find that the total number density of the
strongest \oii-emitting galaxies, i.e., those with $L_{\mathrm{\oiim}}
> 10^{42}~\ergs$, declines by a factor of $\gtrsim2.5$ between
$z\sim1.35$ and $z\sim0.84$.  A linear fit to the four points gives a
slope of $\sim0.9$ dex per unit redshift:  

\begin{equation}
\log\left[\phi(L_{\mathrm{\oiim}}>10^{42}~\ergs)\right] = a~z - b \mathrm{,}
\end{equation}

\noindent with $(a,b)$ = $(0.90\pm0.14, -4.43\pm0.18)$ and
$\phi(L_{\mathrm{\oiim}}>10^{42}~\ergs)$ in $\mathrm{Mpc^{-3}}$.  This
fit is shown as the dashed line in Figure~\ref{fig5}.

%========================
% Figure 5:
%========================

The observed decrease in the number density of \oii-emitting galaxies
may be caused by the decline of the overall number density of galaxies
in the Universe, or by the evolution of the luminosity function
itself.  We attempt to constrain the amount of evolution by measuring
the luminosity at a fixed space density (see, e.g.,
\citealt{brown07}).  We define a characteristic luminosity,
\loiistar, where the space density equals
$10^{-3.5}~\mathrm{dex}^{-1}\mathrm{Mpc}^{-3}$, which has the
advantage that it is independent of the faint end of the LF.  We show
\loiistar{}
as a function of redshift in the bottom panel of Figure~\ref{fig5},
and list the results in Table~\ref{tbl2}.  Once again, the vertical
error bars are dominated by cosmic variance.  We find that \loiistar{}
declines by a factor of $\sim 1.8$ between $z\sim1.35$ and
$z\sim0.84$.  We perform a linear fit to the four points and obtain a
slope of $\sim0.5$ dex per unit redshift:

\begin{equation}
\log \loiistar
= c~z+d \mathrm{,}
\end{equation}
\noindent with $(c,d)$ = $(0.53\pm0.07, 41.79\pm0.09)$ and
$L_{\mathrm{\oiim}}$ in $\ergs$.  The resulting fit is shown as the
dotted line in Figure~\ref{fig5}.

To summarize, we find that the total number density of the strongest
\oii-emitting galaxies has declined by a factor of $\gtrsim 2.5$
between $z \sim 1.35$ and $z \sim 0.84$, when the Universe aged from
$4.6$~Gyr to $6.4$~Gyr.  This decline may be driven by a decline in
the overall number density of galaxies in the Universe, or by a fading
of the \oii{} LF.  Unfortunately, we are unable to establish whether
number-density evolution, luminosity evolution, or a combination of
both is responsible for the observed evolution.  Nevertheless,
  if we assume that the observed change in the \oii{} LF is
  predominantly due to luminosity evolution, that \loii{} is
  proportional to the SFR (see $\S5$), and that the typical dust
  attenuation in galaxies at fixed \oii{} luminosity does not change
  significantly over $0.75<z<1.45$, then this result implies that the
  SFR in galaxies declines by $\sim25\%$ per Gyr during this epoch.
  In the next section we integrate the full \oii{} LF using some
  simple assumptions, and compare our results with other estimates of
  the SFR density at $z\sim1$.

\section{The Evolution of the Star Formation Rate Density}

%=======================
% Figure 6:
%=======================
\begin{figure}
\epsscale{1.3}
\plotone{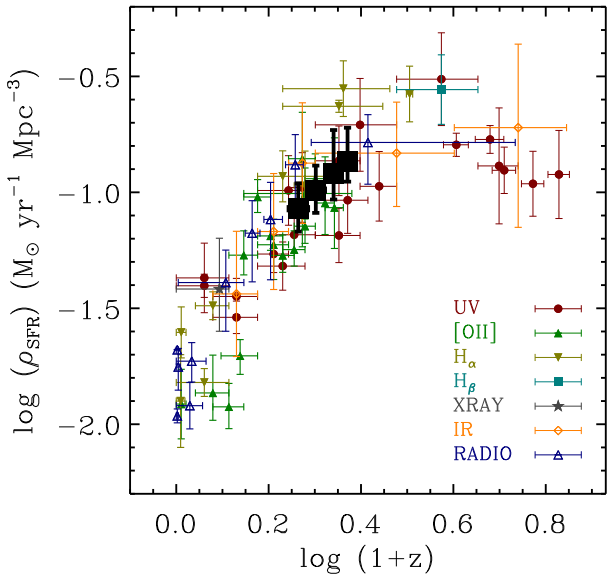}
\caption{SFR density, $\rho_{\rm SFR}$, versus redshift based on
  various multi-wavelength SFR indicators \citep{hopkins04}.  Our four
  estimates of $\rho_{\rm SFR}$ are shown as large filled squares,
  where the vertical error bars are obtained by allowing the turnover
  positions and the slope of the faint end of the \oii{} luminosity
  function to vary in a sensible way (see text for details).}
\label{fig6}
\end{figure}

Converting the observed \oii{} luminosity into a SFR is subject to
considerable random and systematic uncertainties, arising from
variations in dust attenuation, metallicity, and excitation among
star-forming galaxies \citep{kennicutt92, jansen01, kewley04,
  mouhcine05, moustakas06b}.  
Nevertheless, our measurement of
the \oii{} luminosity function affords a valuable opportunity to
constrain the SFR density of the Universe during an important epoch
of cosmic history.

In a recent analysis, \citet{moustakas06b} showed that dust reddening,
as derived using the \ha/\hb{} Balmer decrement, is responsible for
the bulk of the scatter in \oii{} as a SFR indicator, while variations
in metallicity and excitation are second-order effects for most
galaxies.  
Unfortunately, the DEEP2 spectra do not span a
  sufficiently wide wavelength range to include \ha, \hb, and other
  emission-line diagnostics of the metallicity and excitation.
  Therefore, we use the empirical correlation derived by
  \citet{moustakas06b} between the absolute $B$-band magnitude, and
  the \loii/SFR ratio.  This calibration statistically accounts for
  the gross systematic effects of reddening, metallicity, and
  excitation, all of which correlate with optical luminosity, and has
  been shown to work reasonably well for star-forming galaxies at
  $0.7<z<1.4$ \citep{moustakas06b, cooper08}.  Note that an \oii{} SFR
  conversion that is independent of luminosity
  \citep[e.g.,][]{kennicutt98} would severely underestimate the SFR,
  because luminous, star-forming galaxies tend to be dustier and more
  metal-rich \citep{moustakas06b}.

Another potential concern is that some fraction of the \oii{}
  emission might be arising from an active galactic nucleus (AGN)
  rather than star formation.  Although the AGN-sensitive \nii/\ha{}
  ratio \citep{veilleux87, kewley01} lies in the near-infrared at
  $z>0.75$, AGN that contribute significantly to the optical
  emission-line spectrum also can be identified using \oiiilam/\hb.
  In the DEEP2 spectra, \oiii/\hb{} is measurable for galaxies at
  $0.75<z<0.85$, comprising roughly one-third of the sample in our
  lowest redshift bin.  Among these objects, we find $2\%$ ($32/1485$)
  with $\log\,($\oiii/\hb$)>0.7$, indicative of AGN activity.  We can
  also leverage deep X-ray observations of the EGS to identify AGN
  \citep{laird09}.  Among the $2579$ galaxies in this field, we find
  that $1\%$ ($28/2579$) are also X-ray point sources.  These results
  reveal that powerful AGN constitute a negligible fraction of the
  sources in our sample.  However, even if we have significantly
  underestimated the fraction of AGN in our sample, detailed studies
  show that the physical conditions in the narrow-line regions of AGN
  in the local Universe disfavor \oii{} emission, which is one
  advantage of using \oii{} as a SFR tracer \citep{ho05}.  Indeed, we
  will show below that our \oii-based estimate of the SFR density at
  $0.84<z<1.35$ agrees remarkably well with other multi-wavelength
  studies. Therefore, we conclude that AGN contamination is a
negligible source of error on our results.

Before integrating the observed \oii~LF to derive the SFR density,
$\rho_{\rm SFR}$, we must make some assumptions regarding the form of
the faint end of the LF (see $\S3$).  First, we allow the luminosity
of the turnover in the LF in each redshift bin to vary over a sensible
range of values to account for the uncertainties in our completeness.
Specifically, for the two lowest redshift bins, we assume
$\log(L_{\mathrm{TO}})$ = $41.68\pm0.10$ and $41.75\pm0.10$~$\ergs$,
respectively, while for the two highest redshift bins, we adopt a
fainter lower limit: $\log(L_{\mathrm{TO}})$ = $41.93^{+0.10}_{-0.20}$
and $42.00^{+0.10}_{-0.20}$~$\ergs$.

Second, we must assume a form for the \oii~LF fainter than the
turnover luminosity.  Previous studies (e.g. \citealt{gallego02,ly07})
have assumed that the \oii~LF is a Schechter function, which
is a power law at the faint end.  However, the faint-end slope,
$\alpha_{\mathrm{faint}}$, is usually not well-constrained.
Consequently, hereafter we allow $\alpha_{\mathrm{faint}}$ to vary
between $-1.1$ and $-1.5$, which brackets the value,
$\alpha_{\mathrm{faint}}=-1.3\pm0.2$, that we measure from the lowest
redshift bin in Figure~\ref{fig1}. For comparison, \citet{willmer06}
assumed a faint-end slope of $-1.3$ for the $B$-band LF of blue
galaxies at $0.3<z<1.3$.

To summarize, we assume that the \oii~luminosity function is a double
power law with slope:
 $$ \left\{ \begin{array}{lcc} \alpha_{\mathrm{faint}} = -1.3\pm0.2 &
  {\rm for} & L_{\mathrm{\oiim}} < L_{\mathrm{TO}} \\ \alpha &
  {\rm for} & L_{\mathrm{\oiim}} \geq L_{\mathrm{TO}}
        \end{array} \right. $$
\noindent where $\alpha$ is derived from our fit to the bright part of
the LF where we are complete.  Given these assumptions, we integrate
the observed LFs and list the results in Table~\ref{tbl2}.

Instead of using the $B$-band luminosity of each individual galaxy, we
obtain a statistical estimate of $M_B$ for each object from \loii{}
using the thin dashed line in Figure~\ref{fig3}.  We then calculate
the appropriate \loii/SFR conversion factor by interpolating Table~2
in \citet{moustakas06b} to derive the SFR.

Figure~\ref{fig5} compares our measurements of $\rho_{\rm SFR}$ at
$z\sim1$ against a large compilation of multi-wavelength measurements
from the literature by \citet{hopkins04}.  Our results agree
remarkably well with these independent measurements considering the
uncertainties in converting \loii{} into a SFR, and our incompleteness
at the faint end of the LF.

\section{Conclusions}

Because its blue rest-frame wavelength and intrinsic strength allow it
to be measured up to $z\sim1.6$ in the optical, the
\oiidoublet~doublet plays a unique role in the study of galaxy
evolution. We have used spectroscopy of $\sim14,000$ galaxies from the
DEEP2 galaxy redshift survey to measure the \oii~luminosity function
at $0.75 < z < 1.45$. Our sample is orders-of-magnitude larger than
previous spectroscopic studies, over a considerable larger area
spanning four independent fields, allowing us to minimize the
systematic effects of cosmic variance.  Our principal results are
given in Tables~\ref{tbl1} and \ref{tbl2}, and illustrated in
Figures~\ref{fig1} and \ref{fig5}.  We found that the bright part of
the \oii~LF is well-represented by a power law $dN/dL \propto
L^{\alpha}$ with slope $\alpha \sim -3$.  However, survey
incompleteness prevented us from constraining the faint end of the
LF. 

We measured the evolution of the \oii~LF using two quantities that
only rely on the bright part of the \oii{} LF where we are
statistically complete.  First, we calculated the total number density
of galaxies with $L_{\mathrm{\oiim}}>10^{42}~\ergs$, and found that it
has declined by a factor of $\gtrsim 2.5$ between $z\sim1.35$ and
$z\sim0.84$.  Second, we calculated the characteristic luminosity, 
the luminosity where the space density of \oii-emitting galaxies
equals $10^{-3.5}~\mathrm{dex^{-1}Mpc^{-3}}$, and found that it has
declined by a factor of $\sim1.8$ over the same redshift interval.
Assuming that the \oii~luminosity is proportional to the SFR, these
results imply that the SFR in galaxies declined by $\sim 25\%$ per Gyr
during this epoch.

Finally, we used the empirical calibration between \loii{} and SFR
published by \citet{moustakas06b}, and adopted some simple assumptions
regarding the faint end of the \oii{} LF, to obtain an estimate of the
integrated SFR density, $\rho_{\rm SFR}$, in four redshift bins
centered around $z\sim1$.  We found that, despite the considerable
uncertainties, the evolution we measure is consistent with previous
measurements based on a variety of independent, multi-wavelength SFR
indicators.

\acknowledgments 

 It is a pleasure to thank Jeffrey A. Newman, David W. Hogg
  and the anonymous referee for numerous comments that helped improve
  the manuscript. We would also like to thank Benjamin J. Weiner for
  sharing his \oii~measurements for comparison, and for pointing out
  that the variation in the wavelength coverage of the DEEP2
    spectra must be included in our estimate of the completeness
  function. We also thank Chun Ly and Takashi Murayama for kindly
  sharing their \oii~ luminosity function data for comparison.
  The authors acknowledge funding support from NSF grant AST-0607701, 
  NASA grant 06-GALEX06-0030 and \emph{Spitzer} grant G05-AR-50443.

Funding for the DEEP2 survey has been provided by NSF grants
AST95-09298, AST-0071048, AST-0071198, AST-0507428, and AST-0507483 as
well as NASA LTSA grant NNG04GC89G.

Some of the data presented herein were obtained at the W. M. Keck
Observatory, which is operated as a scientific partnership among the
California Institute of Technology, the University of California and
the National Aeronautics and Space Administration. The Observatory was
made possible by the generous financial support of the W. M. Keck
Foundation. The DEEP2 team and Keck Observatory acknowledge the very
significant cultural role and reverence that the summit of Mauna Kea
has always had within the indigenous Hawaiian community and appreciate
the opportunity to conduct observations from this mountain.

\newpage
%=======================
% Table 1
%=======================
\clearpage
\begin{landscape}
\begin{deluxetable}{cccccccccccccccc}
\tabletypesize{\scriptsize}
\tablecolumns{16}
\tablecaption{observed [\ion{O}{2}]~luminosity function}
\tablehead{
\colhead{$\log L_{\mathrm{\oiim}}$}  &  \multicolumn{3}{c}{\underline{$ ~~~0.75 < z < 0.93~~~ $}} &  \colhead{}  &
\multicolumn{3}{c}{\underline{$ ~~~0.93 < z < 1.10~~~ $}} & \colhead{} &
\multicolumn{3}{c}{\underline{$ ~~~1.10 < z < 1.28~~~ $}} &  \colhead{}  &
\multicolumn{3}{c}{\underline{$ ~~~1.28 < z < 1.45~~~ $}} \\
\colhead{(ergs s$^{-1})$} & \colhead{$\Phi$}  & \colhead{$\sigma_{\mathrm{cv}}$} & \colhead{$N_{gal}$}  &
\colhead{}   & \colhead{$\Phi$}   & \colhead{$\sigma_{\mathrm{cv}}$} & \colhead{$N_{gal}$}  & \colhead{} &
\colhead{$\Phi$}   & \colhead{$\sigma_{\mathrm{cv}}$} & \colhead{$N_{gal}$}    &
\colhead{}   & \colhead{$\Phi$}   & \colhead{$\sigma_{\mathrm{cv}}$} & \colhead{$N_{gal}$}}
\startdata
$41.15$ & $49.49^{+3.58}_{-3.34}$ & $ 7.30$ & $553$ &  & \nodata & \nodata & \nodata &  & \nodata & \nodata & \nodata &  & \nodata & \nodata & \nodata \\
$41.30$ & $60.57^{+3.46}_{-3.27}$ & $19.41$ & $918$ &  & $20.69^{+2.47}_{-2.22}$ & $ 7.00$ & $259$ &  & \nodata & \nodata & \nodata &  & \nodata & \nodata & \nodata \\
$41.45$ & $56.07^{+2.24}_{-2.15}$ & $15.88$ & $1095$ &  & $28.25^{+2.03}_{-1.90}$ & $ 2.80$ & $524$ &  & \nodata & \nodata & \nodata &  & \nodata & \nodata & \nodata \\
$41.60$ & $46.93^{+1.94}_{-1.87}$ & $ 7.64$ & $1019$ &  & $35.57^{+1.81}_{-1.72}$ & $ 3.95$ & $779$ &  & $18.86^{+1.97}_{-1.79}$ & $ 1.57$ & $349$ &  & \nodata & \nodata & \nodata \\
$41.75$ & $29.00^{+1.37}_{-1.31}$ & $ 5.82$ & $669$ &  & $31.30^{+1.44}_{-1.38}$ & $ 1.32$ & $779$ &  & $22.30^{+1.15}_{-1.09}$ & $ 2.03$ & $571$ &  & $12.17^{+1.19}_{-1.09}$ & $ 2.22$ & $207$ \\
$41.90$ & $17.68^{+0.94}_{-0.90}$ & $ 3.68$ & $429$ &  & $20.38^{+0.93}_{-0.89}$ & $ 2.64$ & $574$ &  & $23.61^{+1.32}_{-1.25}$ & $ 1.53$ & $658$ &  & $15.53^{+1.18}_{-1.09}$ & $ 1.74$ & $303$ \\
$42.05$ & $ 7.91^{+0.80}_{-0.72}$ & $ 1.57$ & $176$ &  & $11.60^{+0.72}_{-0.68}$ & $ 1.30$ & $332$ &  & $16.60^{+0.93}_{-0.88}$ & $ 0.44$ & $506$ &  & $16.66^{+1.23}_{-1.15}$ & $ 0.94$ & $376$ \\
$42.20$ & $ 4.70^{+0.55}_{-0.50}$ & $ 1.63$ & $104$ &  & $ 6.14^{+0.81}_{-0.72}$ & $ 0.46$ & $150$ &  & $ 9.63^{+0.60}_{-0.57}$ & $ 0.96$ & $321$ &  & $ 9.70^{+0.77}_{-0.71}$ & $ 0.53$ & $247$ \\
$42.35$ & $ 1.62^{+0.35}_{-0.29}$ & $ 0.55$ & $35$ &  & $ 2.77^{+0.60}_{-0.50}$ & $ 0.98$ & $71$ &  & $ 4.52^{+0.42}_{-0.39}$ & $ 1.16$ & $150$ &  & $ 5.59^{+0.51}_{-0.47}$ & $ 0.97$ & $155$ \\
$42.50$ & $ 1.10^{+0.49}_{-0.35}$ & $ 0.44$ & $19$ &  & $ 1.13^{+0.54}_{-0.38}$ & $ 0.34$ & $23$ &  & $ 2.33^{+0.33}_{-0.29}$ & $ 0.56$ & $73$ &  & $ 3.99^{+0.51}_{-0.45}$ & $ 0.38$ & $108$ \\
$42.65$ & $ 0.18^{+0.15}_{-0.09}$ & $ 0.11$ & $4$ &  & $ 0.78^{+0.22}_{-0.17}$ & $ 0.32$ & $21$ &  & $ 1.15^{+0.24}_{-0.20}$ & $ 0.33$ & $37$ &  & $ 1.93^{+0.32}_{-0.28}$ & $ 0.48$ & $52$ \\
$42.80$ & $ 0.18^{+0.21}_{-0.11}$ & $ 0.24$ & $3$ &  & $ 0.27^{+0.14}_{-0.10}$ & $ 0.18$ & $8$ &  & $ 0.27^{+0.12}_{-0.09}$ & $ 0.11$ & $10$ &  & $ 0.87^{+0.23}_{-0.18}$ & $ 0.28$ & $24$ \\
$42.95$ & $ 0.27^{+0.15}_{-0.10}$ & $ 0.12$ & $7$ &  & $ 0.07^{+0.09}_{-0.04}$ & $ 0.06$ & $2$ &  & $ 0.34^{+0.18}_{-0.12}$ & $ 0.21$ & $9$ &  & $ 0.55^{+0.19}_{-0.14}$ & $ 0.18$ & $15$ \\
$43.10$ & $ 0.09^{+0.12}_{-0.06}$ & $ 0.09$ & $2$ &  & $ 0.15^{+0.12}_{-0.07}$ & $ 0.08$ & $4$ &  & $ 0.17^{+0.11}_{-0.07}$ & $ 0.10$ & $6$ &  & $ 0.11^{+0.11}_{-0.06}$ & $ 0.12$ & $3$ \\
$43.25$ & \nodata & \nodata & \nodata &  & $ 0.06^{+0.09}_{-0.04}$ & $ 0.08$ & $2$ &  & $ 0.06^{+0.08}_{-0.04}$ & $ 0.06$ & $2$ &  & $ 0.20^{+0.13}_{-0.08}$ & $ 0.13$ & $6$ \\
$43.40$ & $ 0.04^{+0.08}_{-0.03}$ & $ 0.08$ & $1$ &  & \nodata & \nodata & \nodata &  & $ 0.03^{+0.08}_{-0.03}$ & $ 0.05$ & $1$ &  & $ 0.13^{+0.11}_{-0.06}$ & $ 0.08$ & $4$ \\
$43.55$ & \nodata & \nodata & \nodata &  & $ 0.03^{+0.08}_{-0.03}$ & $ 0.06$ & $1$ &  & $ 0.03^{+0.06}_{-0.02}$ & $ 0.04$ & $1$ &  & $ 0.18^{+0.21}_{-0.11}$ & $ 0.24$ & $3$ \\
$43.70$ & \nodata & \nodata & \nodata &  & \nodata & \nodata & \nodata &  & \nodata & \nodata & \nodata &  & $ 0.03^{+0.07}_{-0.02}$ & $ 0.07$ & $1$ \\
$43.85$ & \nodata & \nodata & \nodata &  & $ 0.03^{+0.08}_{-0.03}$ & $ 0.05$ & $1$ &  & \nodata & \nodata & \nodata &  & $ 0.03^{+0.08}_{-0.03}$ & $ 0.04$ & $1$ \\
\enddata
\tablecomments{
$\Phi$ is in units of $10^{-4}~\mathrm{dex}^{-1}$ Mpc$^{-3}$,
  $\sigma_{\rm cv}$ is the $1\sigma$ uncertainty in $\Phi$ due to
  cosmic variance, and $N_{gal}$ is the number of galaxies in each
  bin.  For reference, the median redshifts of the sources in each of
  the four redshift bins are $0.837$, $1.005$, $1.191$, and $1.349$,
  respectively.
}
\label{tbl1}
\end{deluxetable}
\clearpage

\end{landscape}

%=======================
% Table 2
%=======================
\clearpage
\begin{deluxetable}{ccccc}
\tabletypesize{\small}
\tablecolumns{5}
\tablewidth{0pc}
\tablecaption{parameters of the observed \oii~ luminosity functions}
\tablehead{
\colhead{Quantity}  &  $ 0.75 < z < 0.93 $ &  $ 0.93 < z < 1.10 $ &
$ 1.10 < z < 1.28 $ & $ 1.28 < z < 1.45 $}
\startdata
$z_\mathrm{median}$ & $0.837$ & $1.005$ & $1.191$ & $1.349$ \\
$N_\mathrm{gal}$ & $5471$ & $3771$ & $3018$ & $1684$ \\
Age of Universe ($\mathrm{Gyr}$) & $6.46$ & $5.73$ & $5.06$ & $4.59$ \\
$\alpha$ & $-3.01\pm 0.07$ & $-2.98\pm 0.08$ & $-3.03\pm 0.07$ & $-2.79\pm 0.12$ \\
$\beta$ & $-4.01\pm 0.04$ & $-3.86\pm 0.04$ & $-3.67\pm 0.03$ & $-3.49\pm 0.03$ \\
$\log \Phi(L_{\mathrm{\oiim}}>10^{42}~\ergs)$\tablenotemark{a} & $-3.67^{+ 0.10}_{- 0.13}$ &
 $-3.53^{+ 0.07}_{- 0.08}$ & $-3.32^{+ 0.08}_{- 0.10}$ & $-3.21^{+ 0.01}_{- 0.01}$ \\
$\log$~\loiistar\tablenotemark{b} & $42.24\pm 0.06$ & $42.32\pm 0.04$ & $42.42\pm 0.03$ & $42.50\pm 0.01$ \\
$\log L_{\mathrm{TO}}$\tablenotemark{c} & $41.68\pm0.10$ & $41.75\pm0.10$ & $41.93^{+0.10}_{-0.20}$ & $42.00^{+0.10}_{-2.00}$ \\
$\log~\rho(\mathrm{L_{\oiim}})$\tablenotemark{d} & $39.34^{+ 0.19}_{- 0.16}$ & $39.40^{+ 0.19}_{- 0.16}$ & $39.44^{+ 0.30}_{- 0.17}$ & $39.47^{+ 0.24}_{- 0.13}$ \\
$\rho_{\mathrm SFR}$\tablenotemark{e} & $ 0.09^{+ 0.02}_{- 0.02}$ & $ 0.10^{+ 0.03}_{- 0.02}$ & $ 0.12^{+ 0.07}_{- 0.03}$ & $ 0.14^{+ 0.05}_{- 0.03}$ \\
\enddata
\tablenotetext{a}{$\phi\equiv\int_{L}^{\infty}\Phi(L)\,dL$ is
  in $\mathrm{Mpc^{-3}}$.}
\tablenotetext{b}{The characteristic luminosity, \loiistar, where the space density equals
  $10^{-3.5}\mathrm{dex^{-1}~Mpc^{-3}}$, in $\ergs$.}
\tablenotetext{c}{Luminosity of the turnover in the luminosity
  function, $L_{\mathrm{TO}}$, in $\ergs$.}
\tablenotetext{d}{Integrated luminosity density, $\rho(\mathrm{L_{\oiim}})$, in $\ergs$$\mathrm{~Mpc^{-3}}$.}
\tablenotetext{e}{Star formation rate density in
  $\mathcal{M}_\odot~\mathrm{yr^{-1}~Mpc^{-3}}$.}
\label{tbl2}
\end{deluxetable}
\clearpage


\begin{thebibliography}{}

\bibitem[Blanton(2006)]{blanton06}
  Blanton, M.~R., et al.\ 2006, \apj, 648, 268

\bibitem[Blanton \& Roweis(2007)]{blanton07}
  Blanton, M.~R., \& Roweis, S.\ 2007, \aj, 133, 734

\bibitem[Brown et al.(2007)]{brown07}
  Brown, M.~J.~J., et al.\ 2007, \apj, 654, 858

\bibitem[Coil et al.(2004)]{coil04}
  Coil, A.~L., et al.\ 2004, \apj, 617, 765

\bibitem[Cooper et al.(2008)]{cooper08}
  Cooper, M.~C., et al.\ 2008, \mnras, 383, 1058

\bibitem[Davis et al.(2003)]{davis03}
  Davis, M., et al.\ 2003, \procspie, 4834, 161

\bibitem[Davis et al.(2007)]{davis07}
  Davis, M., et al.\ 2007, \apjl, 660, L1

\bibitem[Faber et al.(2003)]{faber03} 
  Faber, S.~M., et al.\ 2003, \procspie, 4841, 1657 

\bibitem[Felten(1976)]{felten76} 
  Felten, J.~E.\ 1976, \apj, 207, 700 

\bibitem[Gallego et al.(1995)]{gallego95}
  Gallego, J., Zamorano, J., Arag{\'o}n-Salamanca, A., \& Rego, M.\ 1995, \apjl, 455, L1

\bibitem[Gallego et al.(2002)]{gallego02}
  Gallego, J., Garc{\'{\i}}a-Dab{\'o}, C.~E., Zamorano, J., Arag{\'o}n-Salamanca, A., \& Rego, M.\ 2002, \apjl, 570, L1

\bibitem[Gehrels(1986)]{gehrels86} 
  Gehrels, N.\ 1986, \apj, 303, 336 

\bibitem[Glazebrook et al.(1999)]{glazebrook99}
  Glazebrook, K., Blake, C., Economou, F., Lilly, S., \& Colless, M.\ 1999, \mnras, 306, 843

\bibitem[Hammer et al.(1997)]{hammer97}
  Hammer, F., et al.\ 1997, \apj, 481, 49

\bibitem[Hippelein et al.(2003)]{hippelein03}
  Hippelein, H., et al.\ 2003, \aap, 402, 65

\bibitem[Ho(2005)]{ho05}
  Ho, L.~C.\ 2005, \apj, 629, 680 

\bibitem[Hogg et al.(1998)]{hogg98}
  Hogg, D.~W., Cohen, J.~G., Blandford, R., \& Pahre, M.~A.\ 1998, \apj, 504, 622

\bibitem[Hogg(1999)]{hogg99}
  Hogg, D.~W.\ 1999, arXiv:astro-ph/9905116

\bibitem[Hopkins et al.(2000)]{hopkins00} Hopkins, A.~M., 
  Connolly, A.~J., \& Szalay, A.~S.\ 2000, \aj, 120, 2843 

\bibitem[Hopkins(2004)]{hopkins04}
  Hopkins, A.~M.\ 2004, \apj, 504, 622

\bibitem[Hopkins \& Beacom(2006)]{hopkins06}
  Hopkins, A.~M., \& Beacom, J.~F.\ 2006, \apj, 651, 142

\bibitem[Isobe et al.(1990)]{isobe90}
  Isobe, T., Feigelson, E.~D., Akritas, M.~G., \& Babu, G.~J.\ 1990, \apj, 364, 104

\bibitem[Jansen et al.(2001)]{jansen01}
  Jansen, R.~A., Franx, M., \& Fabricant, D.\ 2001, \apj, 551, 825

\bibitem[Kashikawa et al.(2004)]{kashikawa04}
  Kashikawa, N., et al.\ 2004, \pasj, 56, 1011 

\bibitem[Kauffmann et al.(2003)]{kauffmann03}
  Kauffmann, G. et~al.\ 2003, \mnras, 346, 1055

\bibitem[Kennicutt(1983)]{kennicutt83}
  Kennicutt, R.~C.\ 1983, \apj, 272, 54

\bibitem[Kennicutt(1992)]{kennicutt92}
  Kennicutt, R.~C.\ 1992, \apj, 388, 310

\bibitem[Kennicutt(1998)]{kennicutt98}
  Kennicutt, R.~C.\ 1998, \araa, 36, 189

\bibitem[Kennicutt(2008)]{kennicutt08}
  Kennicutt, R.~C.\ 2008, \apjs, 178, 247

\bibitem[Kewley et al.(2001)]{kewley01}
  Kewley, L.~J., Heisler, C.~A., Dopita, M.~A., \& Lumsden, S.\ 2001, \apjs, 132, 37 

\bibitem[Kewley et al.(2004)]{kewley04} 
  Kewley, L.~J., Geller, M.~J., \& Jansen, R.~A.\ 2004, \aj, 127, 2002 

\bibitem[Laird et al.(2009)]{laird09}
  Laird, E.~S., et al.\ 2009, \apjs, 180, 102 

\bibitem[Lilly et al.(1996)]{lilly96}
  Lilly, S.~J., Le Fevre, O., Hammer, F., \& Crampton, D.\ 1996, \apj, 460, L1

\bibitem[Lin et al.(2008)]{lin08}
  Lin, L., et al.\ 2008, \apj, 681, 232

\bibitem[Ly et al.(2007)]{ly07}
  Ly, C., et al.\ 2007, \apj, 657, 738

\bibitem[Madau et al.(1996)]{madau96}
  Madau, P., Ferguson, H. C., Dickinson, M. E., Giavalisco, M.,
  Steidel, C. C., Fruchter, A. 1996, \mnras, 283, 1388

\bibitem[Mouhcine et al.(2005)]{mouhcine05}
  Mouhcine, M., Lewis, I., Jones, B., Lamareille, F., Maddox, S.~J., 
  \& Contini, T.\ 2005, \mnras, 362, 1143 

\bibitem[Moustakas \& Kennicutt(2006)]{moustakas06a} 
  Moustakas, J., \& Kennicutt, R.~C., Jr.\ 2006, \apjs, 164, 81 

\bibitem[Moustakas et al.(2006)]{moustakas06b}
  Moustakas, J., Kennicutt, R.~C., Jr., \& Tremonti, C.~A.\ 2006, \apj, 642, 775

\bibitem[Rigopoulou et al.(2005)]{rigopoulou05}
  Rigopoulou, D.\ 2005, \aap, 440, 61

\bibitem[Schechter(1976)]{schechter76}
  Schechter, P.\ 1976, \apj, 203, 297

\bibitem[Scoville et al.(2007)]{scoville07}
  Scoville, N., et al.\ 2007, \apjs, 172, 150 

\bibitem[Takahashi et al.(2007)]{takahashi07}
  Takahashi, M., et al.\ 2007, \apjs, 172, 456

\bibitem[Taniguchi et al.(2007)]{taniguchi07}
  Taniguchi, Y., et al.\ 2007, \apjs, 172, 9 

\bibitem[Teplitz et al.(2003)]{teplitz03}
  Teplitz, H.~I., Collins, N.~R., Gardner, J.~P., Hill, R.~S., \& Rhodes, J.\ 2003, \apj, 589, 704

\bibitem[Tresse et al.(1998)]{tresse98}
  Tresse, L., \& Maddox, S.~J.\ 1998, \apj, 495, 691

\bibitem[Tresse et al.(2002)]{tresse02}
  Tresse, L., Maddox, S.~J., Le Fevre, O., \& Cuby, J.~-G.\ 2002, \mnras, 337, 369

\bibitem[Veilleux \& Osterbrock(1987)]{veilleux87}
   Veilleux, S., \& Osterbrock, D.~E.\ 1987, \apjs, 63, 295 

\bibitem[Weiner et al.(2007)]{weiner07}
  Weiner, B.~J., et al.\ 2007, \apjl, 660, L39

\bibitem[Willmer et al.(2006)]{willmer06}
  Willmer, C.~N.~A., et al.\ 2006, \apj, 647, 853

\bibitem[Yan et al.(1999)]{yan99}
  Yan, L., McCarthy, P.~J., Freudling, W., Teplitz, H.~I., Malumuth, E.~M., Weymann, R.~J., \& Malkan, M.~a.\ 1999, \apj, 519, L47

\end{thebibliography}
\end{document}